\documentclass[prb,aps,twocolumn,showpacs]{revtex4}
\usepackage{epsfig}
\begin{document}

\title{Anomalous high energy dispersion in photoemission spectra from insulating cuprates}

\author{F. Ronning$^{1,2}$, K. M. Shen$^1$, N. P. Armitage$^{1,\dag}$,
A. Damascelli$^{1,\ddag}$, D.H. Lu$^1$, Z.-X. Shen$^1$, L. L.
Miller$^3$, C. Kim$^{1,4}$}

\affiliation{$^1$Department of Physics, Applied Physics and
Stanford Synchrotron Radiation Laboratory, Stanford University,
Stanford, CA 94305}

\affiliation{$^2$MST-10 Division, Los Alamos National Laboratory,
Los Alamos, NM 87545 }

\affiliation{$^3$Chemistry Department, University of Oregon,
Eugene, OR 97404}

\affiliation{$^4$Institute of Physics and Applied Physics, Yonsei
University, Seoul, Korea}

\date{\today}

\begin{abstract}
Angle resolved photoelectron spectroscopic measurements have been
performed on an insulating cuprate Ca$_2$CuO$_2$Cl$_2$. High
resolution data taken along the $\Gamma$ to ($\pi$,$\pi$) cut show
an additional dispersive feature that merges with the known
dispersion of the lowest binding energy feature, which follows the
usual strongly renormalized dispersion of $\approx 0.35$ eV. This
higher energy part reveals a dispersion that is very close to the
unrenormalized band predicted by band theory. A transfer of
spectral weight from the low energy feature to the high energy
feature is observed as the $\Gamma$ point is approached. By
comparing with theoretical calculations the high energy feature
observed here demonstrates that the incoherent portion of the
spectral function has significant structure in momentum space due
to the presence of various energy scales.\pacs{71.27.+a, 78.20.Bh,
79.60.Bm, 74.72.-h}
\end{abstract}
\maketitle

%Introduction
Since the early days of high temperature superconductivity (HTSC),
it has been suggested, based on the cuprate phase diagram, that
the microscopic origin of HTSC may hinge on how charges move in a
background of strong antiferromagnetic (AF) interactions. Thus, a
natural starting place is to learn how a single hole behaves in a
half-filled AF system. This is precisely what is measured by angle
resolved photoelectron spectroscopy (ARPES) on the half-filled
cuprates. Indeed, such considerations have motivated many ARPES
studies on parent HTSC compounds such as
Sr$_2$CuO$_2$Cl$_2$,\cite{Wells,Kim,LaRosa,Pothuizen}
Ca$_2$CuO$_2$Cl$_2$,\cite{Ronning}
Nd$_2$CuO$_4$,\cite{ArmitagePRL} and lightly doped
La$_2$CuO$_4$.\cite{Yoshida}

While Local Density Approximation (LDA) calculations predict these
systems to be metallic with the width of the lowest filled energy
band $\sim$2 eV,\cite{Novikov} optical measurements indicate they
are insulators with a band gap of $\sim$1.5 eV,\cite{Choi,Perkins}
which results from the strong on-site Coulomb repulsion. ARPES
results have revealed many other important features of insulating
cuprate compounds. Energy Distribution Curve (EDC) analysis has
revealed a broad low-energy charge transfer band
feature,\cite{Wells} which is commonly referred to as a Zhang-Rice
singlet. It has a band width of ~0.35 eV which has been related to
a renormalization by the exchange interaction $J$. The dispersion
has been well described by the $t$-$t'$-$t''$-$J$ model.\cite{Kim}
In addition, results from Ca$_2$CuO$_2$Cl$_2$ show a {\it
d}-wave-like dispersion and a feature in the occupation
probability $n(k)$ that was termed as a `remnant' Fermi
surface.\cite{Ronning}

All these results were obtained with a lower angular (that is,
momentum) resolution of $2^\circ$ compared to what is commonly
used now ($0.25^\circ$), yet made important contributions in
understanding hole motion in the AF background. As has been found
across the phase diagram of the cuprates, the dramatic increase in
photoemission resolution has revealed new features in the data
which were previously unresolved.\cite{AndreaRMP} In this paper,
we present high $k$-resolution data from Ca$_2$CuO$_2$Cl$_2$ which
reveal yet another surprising result. The lowest energy
excitations near ($\pi/2,\pi/2$) remain consistent with the above
description, but at higher binding energies we observe a broad
rapidly dispersing feature which tracks the unrenormalized band
dispersion. We note that there is significant transfer of spectral
weight from the low energy feature to the high energy one as the
$\Gamma$ point is approached. This description is consistent with
several {\it different} numerical
calculations,\cite{Eskes,Tohyama,Hanke,Tremblay,Dahnken} and our
data experimentally confirms this high energy electronic
structure. We will discuss our observations in the context of the
current theoretical understanding.

%Experimental
Ca$_2$CuO$_2$Cl$_2$ single crystals were grown by a flux
method.\cite{Miller} Experiments were performed at beamline V of
the Stanford Synchrotron Radiation Laboratory (SSRL). Samples,
which were oriented {\it ex situ} prior to the experiments by Laue
x-ray diffraction were cleaved {\it in situ} at a base pressure
better than $5\times 10^{-11}$ torr. Spectra were taken with 25.2,
22.4, and 15.5 eV photons, the total energy and angular resolution
was typically 30 meV and $0.25^\circ$, respectively, and no sign
of charging was observed in these samples.

%FIGURE 1
\begin{figure}[t]
\centering \leavevmode \epsfxsize=8.5cm
\epsfbox{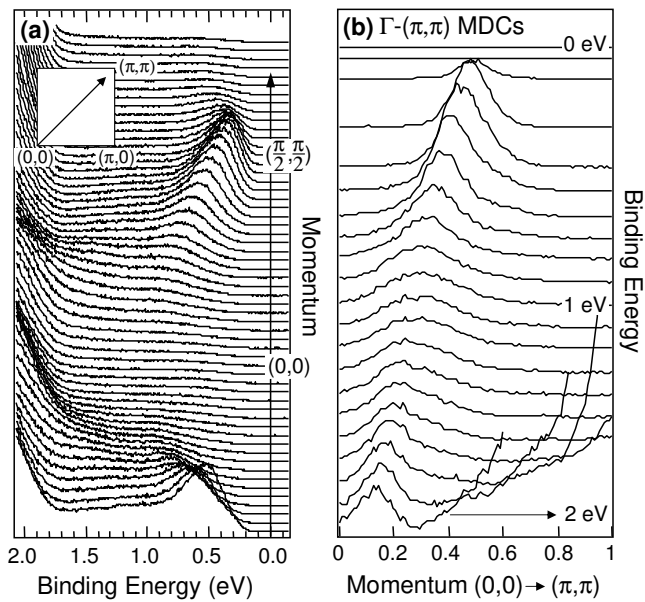} \vspace{.2cm} \caption{ ARPES data
from Ca$_2$CuO$_2$Cl$_2$ along the $\Gamma$-($\pi$,$\pi$)
direction, with 25.2~eV photons. (a) EDC's with the $k$ point for
each curve spaced evenly along the $\Gamma$-($\pi$,$\pi$) line.
(b) MDC curves of the ARPES data shown at equal binding energy
intervals, from 2~eV (bottom) to the chemical potential (top). The
large spectral weight near ($\pi$,$\pi$) for high binding energy
MDCs is due to the O$_{2p_{\pi}}$ states in the main valence
band.}\label{fig1}
\end{figure}

%say this data is compatible with old data
Figure 1(a) shows EDCs of the lowest energy feature along the
nodal direction (that is, $k_x=k_y$) from tetragonal
Ca$_2$CuO$_2$Cl$_2$. It is seen that a broad peak develops near
($\pi/4,\pi/4$), gains spectral weight, and disperses toward the
lower binding energy side as $k$ increases from $\Gamma$ to
($\pi/2,\pi/2$). It then suddenly loses intensity near
($\pi/2,\pi/2$). Furthermore, it leaves no spectral intensity at
the Fermi energy E$_F$, as expected for an insulator. The total
dispersion of the peak is about 0.35 eV. Note also that there is
very little spectral intensity near the $\Gamma$ point. These are
the salient features of the insulating parent HTSC compounds and
have been discussed in the earlier literature.\cite{AndreaRMP}

%new features seen
Our high $k$-resolution data, however, reveal a feature that was
not previously observed. The details can be seen better with an
intensity plot. The data in figure 1 are symmetrized about the
$\Gamma$ point and plotted in Figure 2. The low energy feature
described above is clearly seen between $k=(\pi/4,\pi/4)$ and
$(\pi/2,\pi/2)$ with total dispersion of about 0.35 eV. Examining
the image, one can clearly see that there is a fast dispersing
feature above 0.8 eV binding energy, which is in sharp contrast
with the known and slowly dispersing low energy feature. An
indication of this additional feature was also seen in the earlier
low resolution data roughly 500 meV below the lowest energy
feature whose origin was uncertain, but was attributed to a string
resonance excitation.\cite{Kim} Even though not as clear as in the
Ca$_2$CuO$_2$Cl$_2$ case, we also observed a similar high energy
feature in Sr$_2$CuO$_2$Cl$_2$ (not shown).

%FIGURE 2
\begin{figure}[t]
\centering \leavevmode \epsfxsize=8.5cm
\epsfbox{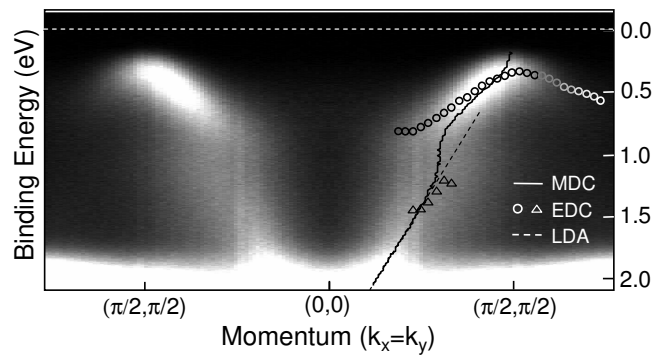} \vspace{.2cm} \caption{Intensity
plot of data shown in figure 1 as functions of the binding energy
and momentum. The data was symmetrized around the $\Gamma$ point.
Also shown on the plot are the dispersions obtained by following
the peak positions of the MDCs (solid line) and the EDCs (circles
and triangles). The results are compared with the shifted
dispersion from the LDA calculation (dashed line).} \label{fig2}
\end{figure}

% MDC dispersion (low energy part & comparison of high E with LDA)
Due to the rapid dispersion this feature is most systematically
characterized by the Momentum Distribution Curves (MDCs) shown in
figure 1(b). The MDC curves are well fit by a Lorentzian up to
binding energies of ~2 eV at which point the main valence band
dominates the spectral weight. The resulting peak positions
determined by fitting the MDCs are overlayed on figure 2 from
which both the low and high energy dispersions discussed above are
evident. One can also track the dispersion from the peak positions
of the EDCs in figure 1(a) which are marked with circles and
triangles. The dispersion obtained from the MDCs and EDCs between
0.3 and 0.7 eV (low energy part) roughly agree. At high energies
($\sim$ 1.3 eV) the MDCs also agree with the very broad and weak
feature seen in the EDCs (triangles). The fact that the MDCs
contain a single peak at all energies suggests that the suppressed
spectral weight near the $\Gamma$ point is not caused by the
matrix element effect, but is due to a significant transfer of
spectral weight from the low energy feature to the rapidly
dispersing high energy feature. Also plotted in the figure is the
dispersion from the LDA calculation shifted by $\sim$ 0.7 eV
(dashed line). Even though there is no independent justification
for shifting the LDA dispersion by 0.7 eV, the agreement between
the slopes of the two dispersions is still remarkable. This
suggests that the rapidly dispersing high energy feature may be
tracking the unrenormalized band.

%FIGURE 3
\begin{figure}[t]
\centering \leavevmode \epsfxsize=8.5cm
\epsfbox{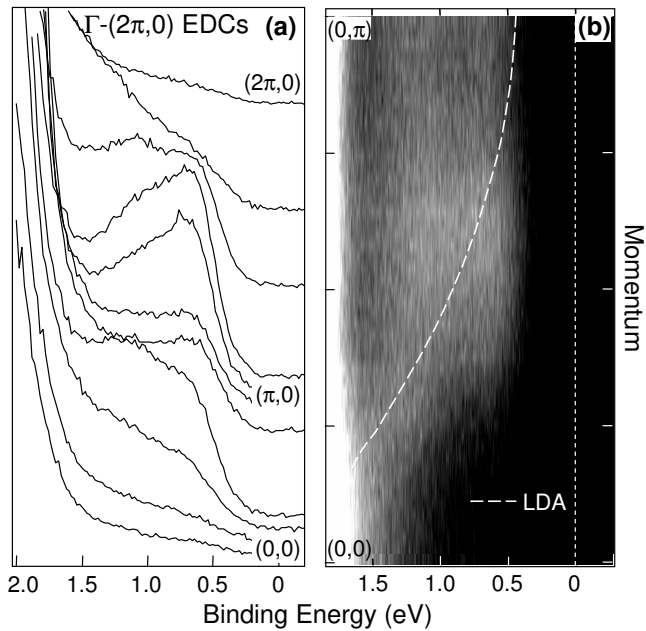} \vspace{.2cm} \caption{(a) EDCs
along the $\Gamma$-($2\pi$,0) direction taken with 22.4~eV
photons. The spectral weight at ($2\pi$,0) is weak, as is the case
for $\Gamma$. (b) Image plot of ARPES data along the
$\Gamma$-(0,$\pi$) cut with 15.5~eV photons. The in-plane
polarization of the light is along the ($\pi$,0) direction.
Overlaid on the plot is the dispersion from LDA calculation
shifted by 0.2 eV (dashed line)}\label{fig3}
\end{figure}

%Kink
The low and high energy dispersive features form a kinklike
feature as can be seen in the image of figure 2. Ignoring the
energy scale, this is very reminiscent of the data on metallic
hole-doped cuprates, where a kink in the quasi-particle dispersion
has been attributed to coupling to a phonon mode of $\sim$
70~meV.\cite{Lanzara} In our data the kinklike feature occurs
roughly 0.5 eV below the top of the valence band, which can be
seen from the MDC derived dispersion in figure 2. However, in the
present case the kink can not originate from coupling to a bosonic
mode. First of all, the system under study is an insulator with a
1.5~eV charge transfer gap and coupling to a bosonic mode is
expected to happen only in metallic systems in the standard
theory. In addition, we also stress that, from the extremely large
widths of the EDC's and MDC's, the features we are studying are
certainly not quasi-particles.\cite{Kim2} In regards to the latter
point, the MDC width of the high energy feature clearly {\it
decreases} as the binding energy increases above 1~eV (see figure
1(b)). This is counterintuitive as the phase space for decay
channels of the photohole should increase as the binding energy
increases, which should thus further raise a red flag when
attempting to interpret these features as quasi-particle bands.

% No weight at Gamma, polarization issue
% Spectral weight transfer explanation for lack of weight at (0,0)
The apparent transfer of spectral weight between the two observed
dispersing features may give an answer to the puzzling observation
of the vanishing spectral weight at the $\Gamma$ point in all
cuprates. Even though the dispersions in cuprates, especially in
insulating compounds, have been said to be renormalized to one
with $\sim$0.3 eV dispersion, the charge transfer peak
mysteriously disappears near the $\Gamma$ point. Traditionally,
this lack of spectral weight at $\Gamma$ has been attributed to
matrix element effects.\cite{Ronning} The photoemission intensity
is roughly expressed as $I\sim |\langle i|{\bf A\cdot p}|f\rangle
|^2$ where $|i\rangle$ and $|f\rangle$ are initial and final
states, respectively. Both the initial $|i\rangle$ (primarily Cu
3$d_{x^2-y^2}$) and final states $|f\rangle$ ($\approx e^{ikz}$)
have even symmetry with respect to the $(\pi,0)$ direction at the
$\Gamma$ point. Since $\bf A\cdot p$ is a dipole operator, it
provides an odd symmetry {\it if} the polarization is in the plane
and along the $(\pi,0)$ direction. This has been the argument for
the vanishing spectral weight at the $\Gamma$ point. In real
cases, however, such symmetry condition is never perfectly met, as
the polarization almost always has an out-of-plane component
(taken here as the $z$-direction).\cite{footnote1} This additional
$z$-component contributes to a non-vanishing matrix element, and
thus the spectral weight should be finite at the $\Gamma$ point.
In fact, the manganites do possess significant spectral weight at
$\Gamma$ due to a band with $d_{3z^2-r^2}$ orbital character for
which the above arguments would be identical.\cite{Dessau}

%(2pi,0) cut
Perhaps even more importantly, we have failed to find any
photoemission intensity at the equivalent $k$-space location of
$(2\pi,0)$ where the selection rule is no longer relevant. This
can be seen in figure 3(a) where upon approaching $(2\pi,0)$ from
$(0,0)$ spectral weight is rapidly pushed to higher binding
energies, and the spectra at $(2\pi,0)$ look identical to $(0,0)$.
Furthermore, note that the matrix element argument dictates that
the entire $(0,0)$ to $(0,\pi)$ cut should have vanishing weight
due to the fact that the in-plane component of polarization of the
light is along the $(\pi,0)$ direction. Although it is indeed
suppressed with respect to the $(0,0)$ to $(\pi,0)$ cut, there
still remains finite low energy weight above $(0,\pi/4)$, as can
be seen in figure 3(b). This not only vanishes upon approaching
the $\Gamma$ point, but also has a high binding energy feature
which disperses rapidly to higher binding energy. This is
identical to the $\Gamma$ to $(\pi,\pi)$ cut, including the
observation that the rapidly dispersing high energy feature has a
similar dispersion to that of LDA, as can be seen in panel 3(b).

All of this suggests that in addition to the matrix elements which
will suppress spectral intensity at the $\Gamma$ point, there is a
transfer of spectral weight to very high binding energies as the
$\Gamma$ point is approached. This is a necessary consequence of
the ARPES sum rules which say that the total spectral weight of
the single particle spectral function must be conserved ($\int
A(k,\omega) d\omega = 1$). Note, this statement is true even for
highly $k$ dependent self energies.

Theoretically, there are several potential explanations for the
multiple features in the photoemission spectra. Cluster
calculations identify the first hole ionization state as a singlet
state and also calculate a large spectrum of excitations which one
will observe in a photoemission experiment.\cite{ZhangRiceTriplet}
While the majority of the states lie too high in energy (such as
the Zhang-Rice triplet state at $\sim$3.5 eV), there are states
which are only $\sim$1.5 eV above the lowest lying singlet state.
Although we can not rule this out conclusively, these states are
still somewhat too large in energy, and there is also no {\it a
priori} reason to expect these states to mimic the dispersion of
the underlying band structure. Thus it is unlikely that they
should correspond to our rapidly dispersing band-like feature.
Alternatively, momentum resolved numerical calculations of the
single band Hubbard model and the extended t-J model have large
amounts of incoherent spectral weight ($>$50$\%$ at many
$k$-points) with significant structure in momentum space spread
over large energies. Surprisingly, although approached differently
from the numerical algorithm and approximations chosen, these
calculated results bear similarities to what we observe in our
data, whether it be from exact diagonalization,\cite{Eskes}
self-consistent Born approximation,\cite{Tohyama} quantum Monte
Carlo,\cite{Hanke} strong coupling perturbation
theory,\cite{Tremblay} or variational cluster perturbation
theory\cite{Dahnken} (a type of generalized dynamical mean field
theory). We also note that the high energy structure was sometimes
referred to as "string" states in the earlier
literature.\cite{DagottoPRB00} The recent calculations show as
$\Gamma$ is approached spectral weight is transferred from a low
energy band that mimics that of the $t$-$J$ model to spectral
weight at higher energies with an energy scale of the hopping
energy $t$, similar to what we see experimentally. The two
features have been interpreted in terms of the coherent motion of
a hole in the antiferromagnetic spin background and the incoherent
motion of a hole hopping within the antiferromagnetic spin
bag.\cite{Hanke} Thus, one might naturally expect the incoherent
feature to track the unrenormalized band.

The outstanding problem with most of the numerical calculations is
that the lowest energy excitation still corresponds to some sort
of quasiparticle like pole, which is not consistent with ARPES
experiments. The very large widths both in EDCs and MDCs are
unphysical lifetimes for an associated quasiparticle. Also, the
low-energy portion of the EDC lineshape can not be fit by any sort
of power-law tail as would be expected from a quasiparticle
picture.\cite{Kyledopingdep} Finally, the doping dependence of the
spectral weight argues that the quasi-particle states which are
formed at the chemical potential in the hole-doped compounds
receive spectral weight from the incoherent background which is
the same feature we identify as the charge transfer band in the
insulator.\cite{Kyledopingdep,RonningNadopedPRB03}

A fully consistent picture of the insulator must therefore also be
able to explain the large widths which are observed. An obvious
candidate for the cause of such large widths is the strong
electron-electron correlation, which suggests that a multiple
state picture be naturally evoked in the photoemission spectra and
that the true quasi-particle has vanishing weight.\cite{Kim2} In
support of such a view, the phase string calculations based on the
$t$-$J$ model have a vanishing quasi-particle residue, and the
spectra, composed entirely of incoherent weight, reasonably
reproduce the experimental lineshapes and
dispersion.\cite{WengPRB} Alternatively, the multiple states that
constitute the broad peaks observed in the experiments may be many
shake-off peaks as in the case of the Franck-Condon
effect.\cite{Kyledopingdep} The appealing aspect of these types of
pictures is that, as we have already demonstrated, our highly
dispersing feature in Ca$_2$CuO$_2$Cl$_2$ can most naturally be
explained as being structure in the incoherent part of the
spectral function. Indeed, in support of the latter view some
calculations predict that the broad peaks still track the $t$-$J$
dispersion even when the polarons are of lattice
origin.\cite{Mishenko,Rosch04} The large widths of the insulator
can then be explained as resulting entirely from the structure in
the incoherent spectral function which tracks both energy scales
in the problem: a `$J$'-like band at low energy and a `$t$'-like
band at high energies.

%Conclusion
In conclusion, we use high resolution ARPES data of
Ca$_2$CuO$_2$Cl$_2$ to elucidate a rapidly dispersing feature at
high binding energies, which is very close to the unrenormalized
bare band. We find a transfer of spectral weight as a function of
momentum from the low energy feature, which tracks the extended
$t$-$J$ model dispersion, to this high energy feature as $\Gamma$
is approached. This interpretation is consistent with several
different calculations working on the half-filled CuO$_2$ plane.
This effect is in addition to the matrix element effects which we
believe are insufficient to explain the lack of spectral weight at
$\Gamma$. The high-energy feature highlights the fact that the
incoherent spectral weight is spread over very large energies and
has significant structure in momentum space. This potentially
allows a previously overlooked aspect of the spectral function to
be used to differentiate various models which purport to
understand the spectra of strongly correlated electron systems.

%Acknowledgement
\begin{acknowledgments} Authors acknowledge helpful discussion with
W. Hanke, A. Mishchenko, D. Senechal, T. Tohyama, A.-M.S.
Tremblay, and Z.-Y. Weng. Authors especially thank T. Tohyama for
sharing unpublished data. F.R. acknowledges support from the
Reines Postdoctoral Fellowship (DOE-LANL). C.K. acknowledges
support from the Korean Science and Engineering Foundation through
the Center for Strongly Correlated Materials Research. The ARPES
measurements at Stanford were also supported by NSF DMR-0304981.
SSRL is operated by the DOE Office of Basic Energy Sciences,
Divisions of Chemical Sciences and Materials Sciences.
\end{acknowledgments}


\begin{thebibliography}{21}
\bibitem[$^\dag$]{Pete}Present address: Dept. de Physique de la Mati\`{e}re Condensée,
Universit\'{e} de Gen\`{e}ve, Switzerland.

\bibitem[$^\ddag$]{Andrea}Present address: Dept. of Physics and Astronomy, University of British
Columbia, Vancouver, Canada, V6T 1Z1.

%Wells' paper
\bibitem{Wells}B.O. Wells, Z.-X. Shen, A.Y. Matsuura, D.M. King, M.A. Kastner,
M. Greven, and R.J. Birgeneau , Phys. Rev. Lett. {\bf 74}, 964
(1995).

%CYKim's 1998 PRL
\bibitem{Kim}C. Kim, P. J. White, Z.-X. Shen, T. Tohyama, Y. Shibata, S. Maekawa,
B. O. Wells, Y. J. Kim, R. J. Birgeneau, and M. A. Kastner , Phys.
Rev. Lett. {\bf 80}, 4245 (1998).

\bibitem{LaRosa}S. LaRosa, I. Vobornik, F. Zwick, H. Berger, M. Grioni,
G. Margaritondo, R.J. Kelley, M. Onellion, and A. Chubukov, Phys.
Rev. B {\bf 56}, R525 (1997).

\bibitem{Pothuizen}J. J. M. Pothuizen, R. Eder, N. T. Hien, M. Matoba, A. A.
Menovsky, and G. A. Sawatzky Phys. Rev. Lett. {\bf 78}, 717
(1997).

%FR's Remnant stuff in Science
\bibitem{Ronning}F. Ronning, C. Kim, D.L. Feng, D.S. Marshall, A.G. Loeser,
L.L. Miller, J.N. Eckstein, I. Bozovic, and Z.-X. Shen, Science
{\bf 282}, 2067 (1998).

%Peter's doping dependence PRL
\bibitem{ArmitagePRL}N.P. Armitage, F. Ronning, D.H. Lu, C. Kim, A. Damascelli,
K.M. Shen, D.L. Feng, H. Eisaki, Z.-X. Shen, P.K. Mang, N. Kaneko,
M. Greven, Y. Onose, Y. Taguchi, Y. Tokura,
Phys. Rev. Lett. {\bf 88}, 257001 (2002).

%Teppei's recent PRL
\bibitem{Yoshida}T. Yoshida, X. J. Zhou, T. Sasagawa, W. L. Yang, P. V. Bogdanov,
A. Lanzara, Z. Hussain, T. Mizokawa, A. Fujimori, H. Eisaki, Z.-X.
Shen, T. Kakeshita, and S. Uchida Phys. Rev. Lett. {\bf 91},
027001 (2003).

%CCOC LDA
\bibitem{Novikov}D. L. Novikov, A. J. Freeman, and J. D. Jorgensen,Phys. Rev. B
{\bf 51}, 6675 (1995).

%IR on SCOC
\bibitem{Choi}H.S. Choi, Y.S. Lee, T.W. Noh, E.J. Choi, Y. Bang, and Y.J. Kim,
Phys. Rev. B {\bf 60}, 4646 (1999).

%IR on SCOC, LCO
\bibitem{Perkins}J. D. Perkins, R. J. Birgeneau, J. M. Graybeal, M. A. Kastner,
and D. S. Kleinberg Phys. Rev. B {\bf 58}, 9390 (1998).

%ZSA paper
%\bibitem{ZSA} ZSA paper.

%Andrea's RMP
\bibitem{AndreaRMP} A. Damascelli, Z.-X. Shen, and Z. Hussain, Rev. Mod. Phys.
{\bf 75}, 473 (2003).

%Lance's SCOC paper
\bibitem{Miller}L. L. Miller, X.L. Wang, S.X. Wang, C. Stassis, D.C. Johnston,
J. Faber, and C.K. Loong, Phys. Rev. B {\bf 41}, 1921 (1990).

%\bibitem{Minimum} The value of the minimum binding energy at
%($\pi/2,\pi/2$) is sample dependent, which results from the pinning
%of the chemical potential inside the gap by various defects which
%vary from sample to sample. The remaining electronic structure
%relative to the binding energy of the charge transfer band at
%($\pi/2,\pi/2$) is indeed sample independent.

%Peter Johnson's Mo paper
%\bibitem{Valla}T. Valla, A. V. Fedorov, P. D. Johnson, and S. L. Hulbert Phys.
%Rev. Lett. 83, 2085-2088 (1999).

%Anton's SRO 327 paper
%\bibitem{Puchkov} A. V. Puchkov, M. C. Schabel, D. N. Basov, T. Startseva, G. Cao,
%T. Timusk, and Z.-X. Shen, Phys. Rev. Lett. {\bf 81}, 2747 (1998).

%Anton's SRO 214 paper
%\bibitem{PuchkovSr214} A. V. Puchkov, Z.-X. Shen, T. Kimura, Y. Tokura,
%Phys. Rev. B {\bf 58}, R13322 (1998).

%Ale's Nature
\bibitem{Lanzara}A. Lanzara, P.V. Bogdanov, X.J. Zhou, S.A. Kellar, D.L.
Feng, E.D. Lu, T. Yoshida, H. Eisaki, A. Fujimori, K. Kishio,
J.-I. Shimoyama, T. Nodak, S. Uchida, Z. Hussain, and Z.-X. Shen,
Nature {\bf 412}, 510 (2001).

%CYKim's PRB
\bibitem{Kim2}C. Kim, F. Ronning, A. Damascelli, D.L. Feng, Z.-X.
Shen, B.O. Wells, Y.J. Kim, R.J. Birgeneau, M.A. Kastner, L. L.
Miller, H. Eisaki, and S. Uchida, Phys. Rev. B {\bf 65}, 174516
(2002).

\bibitem{footnote1}In the current setup the incident light hits the sample
at $\sim$50$^\circ$ from the normal.

%Dan Dessau's PRL
\bibitem{Dessau}D. S. Dessau, T. Saitoh, C.-H. Park, Z.-X. Shen, P. Villella,
N. Hamada, Y. Moritomo, and Y. Tokura, Phys. Rev. Lett. {\bf 81},
192 (1998).

% Energy position of ZRT
\bibitem{ZhangRiceTriplet}H. Eskes and G.A.
Sawatzky, Phys. Rev. B {\bf 44}, 9656 (1991); H. Eskes, L.H. Tjeng, and G.A.
Sawatzky, Phys. Rev. B {\bf 41}, 288 (1990).

%XAS results
%\bibitem{XAS} C. T. Chen {\it et al.}, Phys. Rev. Lett. {\bf 68}, 2543
%(1992). This result is for materials with apical oxygens. Due to
%the ionicity of Cl, it is expected to be even higher for CCOC.

%Elbio's string resonance
%\bibitem{StringResonance}E. Dagotto, Rev. Mod. Phys. {\bf 66}, 763 (1994) and
%references therein.

\bibitem{Eskes}H. Eskes and R. Eder, Phys. Rev. B {\bf 54}, 14226 (1996).

\bibitem{Tohyama}Y. Shibata, T. Tohyama, and S. Maekawa, Phys.
Rev. B, {\bf 59} 1840 (1999).

\bibitem{Hanke}C. Grober, R. Eder, and W. Hanke, Phys. Rev. B {\bf 62}, 4336 (2000).

\bibitem{Tremblay} S. Pairault, D. Senechal, and A.-M.S. Tremblay,
Eur. Phys. J. B, {\bf 16} 85 (2000).

\bibitem{Dahnken}C. Dahnken, M. Aichhorn, W. Hanke,
E. Arrigoni, and M. Potthoff, Phys. Rev. B, {\bf 70}, 245110
(2004).

%\bibitem{Srivistava} P. Srivastava and A. Singh, cond-mat/0402487 (2004).

\bibitem{DagottoPRB00} E. Dagatto, R. Joynt, A. Moreo, S. Bacci,
and E. Gagliano, Phys. Rev. B, {\bf 41}, 9049 (1990).

\bibitem{Kyledopingdep} K.M. Shen, F. Ronning, D.H. Lu, W.S. Lee,
N.J.C. Ingle, W. Meevasana, F. Baumberger, A. Damascelli, N.P.
Armitage, L.L. Miller, Y. Kohsaka, M. Azuma, M. Takano, H. Takagi,
and Z.-X. Shen, Phys. Rev. Lett. {\bf 93}, 267002 (2004).

%\bibitem{fronningPRB03}F. Ronning, C. Kim, K.M. Shen, N.P. Armitage,
%A. Damascelli, D.H. Lu, D.L. Feng, Z.-X. Shen, L.L. Miller, Y.-J.
%Kim, F. Chou, I. Terasaki, Phys. Rev. B, {\bf 67}, 035113 (2003).

%Filip's PRB on Na-doped CCOC
\bibitem{RonningNadopedPRB03} F. Ronning, T. Sasagawa, Y. Kohsaka, K.M. Shen,
A. Damascelli, C.Kim, T.Yoshida, N.P. Armitage, D.H. Lu, D.L.Feng,
L.L. Miller, H.Takagi, and Z.-X. Shen, Phys. Rev. B, {\bf 67},
165101 (2003).

\bibitem{WengPRB} Z. Y. Weng, V. N. Muthukumar, D. N. Sheng, and C. S.
Ting, Phys. Rev. B {\bf 63}, 075102 (2001).

\bibitem{KyleFCBinsulator} K.M. Shen, F. Ronning, W. Meevasana, D.H. Lu,
N.J.C. Ingle, F. Baumberger, W.S. Lee, L.L. Miller, Y. Kohsaka, M.
Azuma, M. Takano, H. Takagi, and Z.-X. Shen, unpublished (2004).

\bibitem{Mishenko} A. Mishchenko and N. Nagaosa, Phys. Rev. Lett. {\bf 93},
036402 (2004).

\bibitem{Rosch04} O. R\"{o}sch and O. Gunnarsson, Eur. Phys. J. B,
{\bf 43}, 11 (2005)

\end{thebibliography}
\end{document}